\documentclass[aps,superscriptaddress,twocolumn,pre]{revtex4-1}
\usepackage{graphicx}
\usepackage{color}
\usepackage{bm}
\usepackage{amsmath,amssymb}
\usepackage[normalem]{ulem}
\usepackage{subfigure}
\usepackage{hyperref}
\usepackage{ifpdf}
\usepackage{color}
\usepackage{xcolor}
\usepackage{multirow}
\usepackage{afterpage}
\usepackage{cancel}
\begin{document}

\title[Chaos in BEC]{Chaotic dynamics of Bose-Einstein condensates in a tilted optical lattice}

\author{C. Selvaraju}
\email{kcselva2020@gmail.com}
\affiliation{Department of Physics, V.S.S Government Arts College (Affiliated to Alagappa University, Karaikudi-630 003, India), Pulankurichi, Tamilnadu 630 405, India}

\author{S. Sabari}
\email{ssabari01@gmail.com}
\affiliation{Instituto de F\'{\i}sica Te\'{o}rica, UNESP--Universidade Estadual Paulista, 01140-070 S\~{a}o Paulo, Brazil}

\author{O.T. Lekeufack}
\email{lekeufackolivier@gmail.com}
\affiliation{Department of Physics, Faculty of Science, University of Douala, P.O. Box 24157 Douala, Cameroon}

\author{F. Kenmogne}
\affiliation{Department of Civil Engineering, Advanced Teacher's Training College of the Technical Education, University of Douala, P.O. Box 1872 Douala-Cameroon}

\author{N. Athavan}
\affiliation{Department of Physics, H.H. The Rajah’s College (Autonomous) (Affiliated to Bharathidasan University, Tiruchirappalli-620024, India), Pudukkottai, Tamilnadu 622001, India}

\begin{abstract}
This study investigates the emergence of chaotic dynamics in Bose-Einstein condensates (BECs) subjected to both alternating (AC) and constant (DC) components of the interaction strength, modeled through the scattering length. We systematically explore how the interplay of AC and DC nonlinearities affect the dynamical evolution of the condensate under a tilted optical lattice potential. Various types of chaos are identified across different parametric regimes, with numerical simulations revealing a clear distinction between regular and chaotic domains. The width of the regular domains is quantified, and the influence of AC and DC components in promoting stochastic behavior is highlighted. Lyapunov exponents, Poincaré sections, and other chaos indicators then confirm the transition to chaotic dynamics, in agreement with analytical expectations. A qualitative conjecture is proposed for the role of these interactions in BEC stabilization. Our findings offer insights into the dynamic control of BECs, with potential applications in quantum simulation and coherent matter-wave engineering, in line with entanglement and quantum transport, that are crucial for developing robust and reliable quantum technologies.
		
\end{abstract}

\date{\today}
\maketitle

\newpage

\section{Introduction}
\label{sec:int}
Bose-Einstein condensates (BECs) are a unique state of matter that have attracted a lot of attention in the scientific community due to their unusual properties and potential applications~\cite{Anderson1995,Bradley1995,Davis1995}. Interactions between particles in a BEC are described by linear and often nonlinear dynamical equations, which can create interesting patterns like vortices and solitons~\cite{Kevrekidis2003,2017Sabarivor,2019Tamilvor,2024Sabarivor,2024Tomiovor,Sabari2025,Komineas2003,Nkenfack2025,Nkenfack2025b}. These dynamics are not always evident to predict and control, but they offer valuable insights into the behavior of complex systems.

In the ultracold regime where temperatures are much smaller than the critical temperature for condensation, a Bose gas may obey the $T=0$ formalism. Most experimental findings in BECs are reproduced and described by the theoretical model based on the nonlinear mean-field GP equation with two-body interactions. The effects of the interatomic interaction lead to a nonlinear term in the GP equation, which is proportional to the \textit{s}-wave scattering length, $a_s$, and the condensate's density~\cite{Dalfovo1999}. One can change the sign and strength of the scattering length by using Feshbach resonance techniques~\cite{Feshbach}. In the 1D homogeneous limit, the GP equation takes the form of a nonlinear Schr\"{o}dinger equation that supports a spectrum of exact soliton solutions. Experiments approach this mathematically ideal scenario by confining the condensate in an elongated and prolate trap, typically with tight radial confinement. However, this quasi-1D geometry is usually accompanied by the presence of weak axial harmonic trapping, which removes the integrability of the system~\cite{Dalfovo1999}.
It is well understood that at low densities, where the inter-atomic distances are significantly greater than the distance scale of atom-atom interactions, the two-body interaction can be described by a scattering length with the effects of higher-order interactions being neglected~\cite{2010Sabari,2017Sabari,2022Sabari,Sabari2022b,Sasireka2025a,Sasireka2025b}.
But, in some experiments, the density of the BECs is considerably high. Particularly, the  progress with BECs on the surface of atomic chips and in atomic waveguides involves a strong compression of the traps and it can appreciably enhance the density of the BECs. Consequently, the simple GP equation (with two-body interaction alone) becomes less convenient. Thus, the dynamics of the BEC needs better description of atom-atom interaction. Such a system comprises three-body interactions. 
Also, the three-body interaction plays an important role in the cases of higher densities and Efimov resonance. Moreover, in a near-collapse situation, not only the nonlocality of the two-body collisions but also three-body interactions should be taken into account~\cite{2010Sabari}.
This system may be an ideal candidate to test the correlations between light and mesoscopic objects, to understand the underlying physics, and to predict the possible applications in quantum information processing experiments which have been recently conducted with ultra-cold bosons in optical resonators~\cite{Byrnes2012,Pelegri2018}.

In recent past years, there have been a lot of works in a different context of BEC with DC and AC parts of the nonlinear interactions~\cite{2018Sabari,2018Tamil,Tamil2019}. Towards the aim of stabilization, Saito and Ueda~\cite{Saito2003}, inspired by the physics of reverted pendulum and suggested a scheme to stabilize 2D solitons with periodically varying time-dependent nonlinearity~\cite{Syu2020}. 
	Thus, periodically oscillating nonlinearity must be needed to balance the repulsive or attractive force to stabilize the BEC~\cite{Lekeufack2024,Sabari2015a,Sasireka2023}. 

	From the nonlinear nature of their dynamics, one of the most interesting features of such BECs is the ability of exhibiting chaotic behavior. Chaos refers to a type of behavior that appears random and unpredictable but is actually governed by underlying laws and can be described mathematically by nonlinear equations~\cite{Gutzwiller1990,Haake1990}. Chaotic systems are highly sensitive to initial conditions, meaning that even small changes in the starting state can lead to vastly different outcomes over time.
In a BEC, chaos often arises due to the complex interactions between the particles. When the BEC is perturbed in some way, such as by introducing an external field or by slightly altering the temperature, the particles can respond in a highly nonlinear and unpredictable manner. Chaos in BECs is closely related to the instability and collapse of the system \cite{Syu2020,Lekeufack2020,Franzosi2003}. This has extensively been studied in recent years, both theoretically and experimentally~\cite{Smerzi1997a,Smerzi1997b,Kagan1997,Uzar2023,Chen2022}. Researchers have found that chaotic behavior can be both a source of frustration and a valuable tool for exploring the properties of these unique quantum systems. Researches in this field have important scientific significance: by understanding the underlying dynamics of chaos in BECs, scientists hope to gain a deeper understanding of the fundamental laws of nature that govern the behavior of matter at the quantum level. The study of chaos in BEC is important for understanding the fundamental nature of quantum mechanics and for developing new technologies such as quantum computing~\cite{Byrnes2012}, control of quantum errors, quantum transport and entanglement \cite{Furuya1998,Song2001,ZhiXiaa2010}, and  quantum sensors~\cite{Pelegri2018}. It also has implications for other areas of physics, such as superfluidity~\cite{Nemirovskii2018} and turbulence~\cite{Zhang2018}.

There are several types of chaos that can arise in dynamical systems, which are systems that evolve over time according to a set of equations. Here are some of the most common types of chaos:

\begin{itemize}
	\item {\bf Periodic chaos:} In a periodic chaotic system, the behavior repeats itself over time, but the repetition is not exact. This means that the system appears to be random, but there is actually some underlying order to the behavior.
	
	\item {\bf Strange attractors:} In a strange attractor, the system evolves towards a complex, fractal-like pattern that is highly sensitive to initial conditions. The attractor can take on a wide range of shapes, depending on the parameters of the system.
	
	\item {\bf Intermittent chaos:} In intermittent chaos, the system alternates between periods of chaotic behavior and periods of regular behavior. This can create a pattern of bursts and lulls in the system's output.
	
	\item {\bf Chaotic transients:} Chaotic transients refer to the initial behavior of a chaotic system before it settles into a more stable pattern. During this transient period, the system may exhibit highly unpredictable and irregular behavior.
	
	\item {\bf Spatiotemporal chaos:} In spatiotemporal chaos, the system evolves over both time and space, creating complex patterns that are difficult to predict or control. This can arise in systems such as fluid dynamics or neural networks.
\end{itemize}

Studying the chaotic behavior of BEC systems, especially under OL potential has revealed increasing improvements both for the theoretical and experimental research communities as to prevent and understand still unpredictable phenomena and applications.
However, as far as we know, there is no work on the chaos in BEC with AC and DC parts of the nonlinear interactions under tilted OL potential. So in the present study, we are interested to study the effects of the two parts of the nonlinear interactions on the chaos in BEC. 
	The novelty of this work lies on the comprehensive analysis of chaotic behavior in BECs with tunable nonlinear interactions that include both constant (DC) and oscillatory (AC) components for two- and three-body terms, placed under a tilted optical lattice potential. Unlike previous studies that typically consider either two-body interactions or time-independent nonlinearities, we present a unified framework incorporating both temporal and spatial modulations of interaction strength. This allows us to identify and classify multiple types of chaos—ranging from small to global—across distinct parameter regimes, offering a richer dynamical picture. From an experimental perspective, our study is motivated by recent advances in manipulating scattering lengths via Feshbach resonance techniques and engineering external potentials in optical lattices. Understanding how interaction modulations induce or suppress chaos is essential for maintaining BEC stability and coherence, which are critical for applications in quantum simulation, quantum computation, and high-precision quantum sensing. The results presented here provide actionable insights for optimizing trapping configurations to avoid unwanted chaotic behavior in practical BEC-based devices.	

The paper will be organized as follows: in section $2$, we draw the model under study and decouple subsequent set of first-order ordinary differential equations ready for the numerical analysis presented in section $3$; section $4$ is devoted to results of analytical and numerical integrations for various combinations of AC and DC impact on the dynamics. 
	In Section $5$, the chaotic dynamics has been quantified with the aid of maximal Lyapunov exponent. In section $6$ a conclusion is drawn to recall the main results of our study and to raise some discussions.

\section{First model}\label{sec:gpe}

We consider a BEC immersed in a tilted optical lattice potential. At very low temperatures, BECs with two- and three-body interactions can be described by the following GP equation \cite{Sabari2022b,Sasireka2025a,Lekeufack2020,2018Sabari,2018Tamil,Tamil2019,2020Sabari}

\begin{figure}[h!]
\centering
\includegraphics[width=0.45\textwidth]{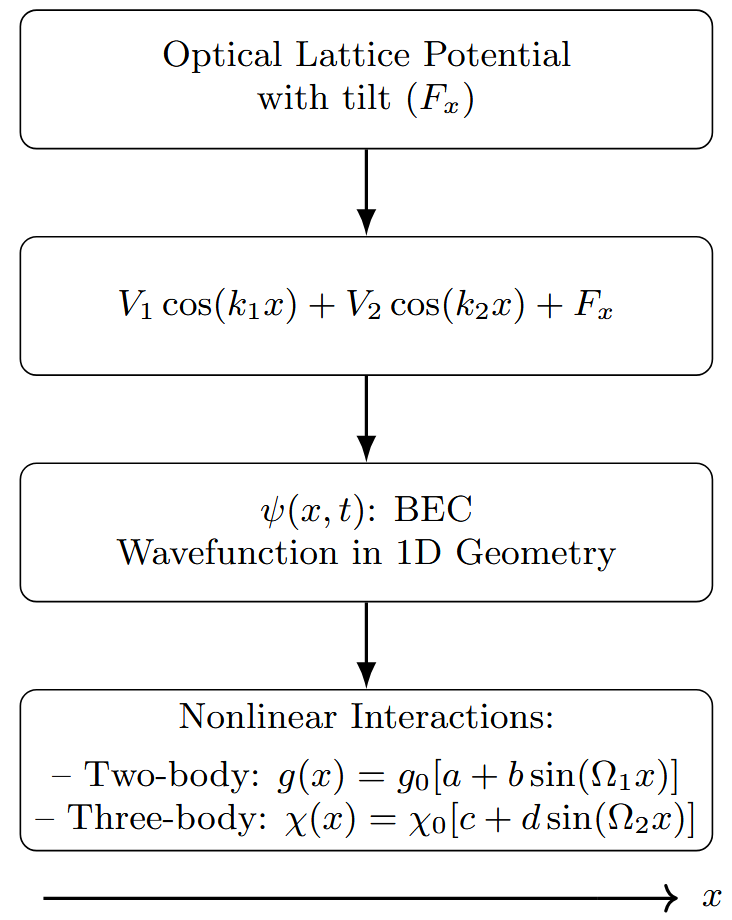}
\caption{Schematic representation of the BEC system under a tilted optical lattice potential with spatially modulated two-body and three-body interactions.}
	\label{fig:bec_schematic}
\end{figure}

\begin{align}
	\mathrm{i}\hbar&\frac{\partial \psi(\mathbf{r},t)}{\partial t} =\left[ -\frac{\hbar^2}{2m}\nabla^2+ \frac{1}{2}m\left(\omega_\perp^2 \rho^2+\omega_x^2 x^2\right)\right]\psi(\mathbf{r},t)\notag \\& + \left(V_{OL}(x)+g(x) |\psi(\mathbf{r},t)|^2 +\chi(x) |\psi(\mathbf{r},t)|^4\right) \psi(\mathbf{r},t)\,\,\, \label{equ:1}
\end{align}
where $\hbar$ is the reduced Planck’s constant, $m$ is the mass of the boson,  $\omega_{\perp}$ and $\omega_{x}$, respectively, are the radial and longitudinal frequencies of the anisotropic trap ($\omega_{\perp} \neq \omega_{x}$) and $\rho^2=y^2+z^2$ denotes the radial distance. The tilted OL potential is applied only in the axial direction, such as to have 
\begin{equation}
	V_{OL}(x) = V_1 cos(\kappa_1 x)+V_1 cos(\kappa_2 x) + Fx, 
\end{equation}

with $V_1$, $V_2$ the amplitudes of the OL potential and $\kappa_1$, $\kappa_2$ are the wave vectors of the laser beams used to create the OL potential. The tilt of the OL potential is described by the term $Fx$, where $F$ denotes the tilt coefficient of the OL. Then, $g(x)$ and $\chi(x)$ are the strengths of the two- and three-body interactions, respectively.

The radial motion can strongly be confined by making the radial trapping frequency $\omega_{\perp}$ much larger than the axial frequency $\omega_{x}$. In this case, the condensate is cigar-shaped, and owing to that, one can take
\begin{equation}
	\psi(\mathbf{r},t)=\phi_0(\rho)\phi(x,t), 
\end{equation} 
where $\phi_0$ is the ground state given by $\phi_0=\sqrt{\frac{1}{\pi a_{\perp}^2} } exp (\frac{\rho^2}{2a_{\perp}^2})$,
with {$a_\perp = \sqrt{\hbar/(m\omega_\perp)}$,
accounting for the radial equation
\begin{align}
	-\frac{\hbar^2}{2m} \nabla_{\rho}^2\phi_0+\frac{m}{2}\omega_{\perp}^2\rho^2\phi_0=\hbar\omega_{\perp}\phi_0.
\end{align}

	To perform the dimensional reduction from three to one spatial dimension, we assume a strong harmonic confinement in the transverse directions ($y$ and $z$), leading to a quasi-1D geometry. This justifies the effective reduction under the Gaussian ansatz.
Now, multiplying both sides of the GP Eq.(\ref{equ:1}) by $\phi_0^*$ and integrating over the radial variable $\rho$, we obtain a quasi-one dimensional (1D) GP equation in its convenient form that reads:

\begin{align}
	\mathrm{i}\hbar\frac{\partial \phi(x,t)}{\partial t} &=\left(-\frac{\hbar^2}{2m}\nabla_x^2+ \frac{1}{2}m \omega_x^2 x^2+V_{OL}(x)\right)\phi(x,t)\notag \\& + \left(\tilde{g}(x) |\phi|^2 +\tilde{\chi}(x)|\phi|^4\right) \phi(x,t)\,\,\, \label{equ:3}
\end{align}
where, $\tilde{g}(x)=g(x)/(2\pi a_{\perp}^2)$ and $\tilde{\chi}(x)=\chi(x)/(3\pi^2a_{\perp}^4)$. It is more convenient to use the above equation (\ref{equ:3}) in a dimensionless form. For this purpose we make the transformation of variables, then, we can get the following normalized 1D GP equation in the absence of harmonic potential:
\begin{align}
	\mathrm{i}\frac{\partial \psi(x,t)}{\partial t} &=\left(-\frac{\partial^2}{\partial x^2}+g\prime(x) |\psi|^2 +\chi\prime(x) |\psi|^4 \right)\psi(x,t)\notag \\& + \left( V\prime_{1}cos(k\prime_1\,x)+V\prime_{2}cos(k\prime_2\,x)+F\prime x\right) \psi(x,t)\,\,\, \label{equ:4}
\end{align}
where, $g\prime(x)= g_0\prime\left[a+b\, sin(\Omega_1\, x)\right]$ and $\chi\prime (x)= \chi_0\prime\left[c+d\, sin(\Omega_2\, x)\right]$.
In order to obtain a simple description and hence a better understanding of the BEC dynamics, we consider $\psi(x,t)$  in the form~\cite{Uzar2023}:

\begin{align}
	\psi(x,t)=\phi(x)e^{(\frac{-i\mu t}{\hbar})}
	\label{equ:5}
\end{align}
where $\mu$ is the chemical potential of the condensate and $\phi(x)$ is a real and normalized wave function, where normalization of the wave function is described as  $\int \phi(x)^2 dx=N$. Substitution of Eq.(\ref{equ:5}) in Eq.(\ref{equ:4}), (after removing the $\prime$) yields the following equation

\begin{align}
	\mu\, \phi(x) &=\left(-\frac{\partial^2}{\partial x^2}+g(x) |\phi(x)|^2 +\chi(x) |\phi(x)|^4 \right)\phi(x)\notag \\& + \left( V_{1}cos(k_1\,x)+V_{2}cos(k_2\,x)+F x\right) \phi(x)\,\,\, \label{equ:6}
\end{align}
\begin{table}[h]
	\centering
	\caption{List of symbols and parameters used in the model}
	\begin{tabular}{ll}
		\hline
		Symbol & \, Meaning \\
		\hline
		$a_\perp$ & Transverse oscillator length, $\sqrt{\hbar/(m\omega_\perp)}$ \\
		$g_2$ & Effective two-body interaction strength \\
		$g_3$ & Effective three-body interaction strength \\
		$F$ & Tilt strength (energy gradient) \\
		$\omega_\perp$ & Transverse trapping frequency \\
		$\mu$ & Chemical potential \\
		\hline
	\end{tabular}
\end{table}

\section{Numerical Analysis}\label{sec:Num}

The above ordinary differential equation is the stationary state GP equation in the case of zero trivial phase which describes the dynamics of the BEC. It is difficult to obtain the exact solution of Eq.(\ref{equ:6}) because of its complexity. Nevertheless, due to its nonlinear aspect, analytical assumptions are made over the dynamics that predict various situations ranging from regular to chaotic behaviors. Due to the non-integrability aspect of the flow with respect to local variables, numerical integration is necessary here.
For that, we need to reduce the system to the first order through the transformation
\begin{subequations}
	\begin{align}
		\frac{d\phi(x)}{dx}&=  y(x) \\
		\frac{d\,y(x)}{dx}&=\left(g(x) |\phi(x)|^2 +\chi(x) |\phi(x)|^4 -\mu \right)\phi(x)\notag \\& + \left( V_{1}cos(k_1\,x)+V_{2}cos(k_2\,x)+F x\right) \phi(x)
	\end{align}\label{equ:7}
\end{subequations}

where,
\begin{subequations}
	\begin{align}
		g(x) = g_0\left[a+b\, sin(\Omega_1\, x)\right]\\
		\chi(x)= \chi_0\left[c+d\, sin(\Omega_2\, x)\right]. 
		\label{equ:51}
	\end{align}
\end{subequations}
	The resulting system of coupled first-order ordinary differential equations (ODEs) was numerically integrated using a fourth-order Runge-Kutta method with adaptive step size control to ensure accuracy and stability.
Next, the above two coupled equations can be solved simultaneously with suitable initial conditions and other system parameters. We kept $k_1=k_2=1$, $\Omega_1=\Omega_2=1$ throughout the paper.

The definition and observation of chaotic behavior in such systems are familiar and well understood~\cite{Lekeufack2008,Tabor1989}. Chaotic systems are those that exhibit a sensitive dependence on initial conditions, meaning that small differences in the initial state of the system can lead to large differences in its behavior over time. In this section, we discuss the occurrence of chaos in BECs with different possible cases.
Unfortunately, unlike an analytical relation from which one can discuss the appearance of chaos for different initial conditions, numerical integration has the drawback of requiring a discrete variation of a parameter that controls the system. Consequently, numerically studying such a semi-quantum system for several values of its control parameter, which may vary within intervals of relatively long length, will demand a cumbersome quantity of plots. Despite our great attention focused only on the influence of the control parameter, the numerical description involves quite a large number of plots. The fourth-order Runge-Kutta algorithm is the scheme we have used. The time step is fixed at $\Delta t=0.005$ together with convenient initial configurations. Generally, we did not explore the full set of all initial conditions, which would require much more extensive numerical calculations and that is beyond the scope of the present paper.  We have used four different indicators, namely:

\begin{enumerate}
	\item [(i)] Trajectory plot which is plotted for $\phi(x)$ vs $\phi\prime(x)$
	
	\item [(ii)] Poincaré surface of section plot which is plotted for $\phi(x)$ vs $\phi\prime(x)$
	
	\item [(iii)] Potential plot which is plotted for $x$ vs $V(x)$
	
	\item [(iv)] Spacial evaluation of wavefunction plot which is plotted for $x$ vs $\phi(x)$
\end{enumerate}

The trajectory plots in chaos theory are often referred to as phase-space plots, namely phase portraits~\cite{Lekeufack2013}. Phase space plots are particularly useful for analyzing chaotic systems because they provide a visual representation of the system's behavior. In a phase space plot, the state of a system is represented by a point in a multidimensional space, with each dimension of the space standing for a different aspect of the system's state. Phase portraits are then a powerful tool for studying chaotic systems in a visual and intuitive way~\cite{Lekeufack2013}. In addition to revealing the behavior of chaotic systems in the phase space, it is possible to make predictions about its future behavior, including whether it will remain in a stable state or undergo fluctuations. Phase portraits may be sufficient to state whether the dynamic is regular or not. Nevertheless, they are not practical when the phase space is of dimension greater than two. Moreover, we cannot easily distinguish roughly between chaotic states and some quasiperiodic ones using only phase portraits.
	Thus, the Poincaré surface of sections are appropriate here. 
The Poincaré surface of sections enables us to characterize the long time dynamics of our model under slight perturbation of initial conditions, as well. They are useful to determine, in particular, the periodicity of the systems evolution~\cite{Lekeufack2008}. Smooth closed curves always determine regular periodic motion, while strange attractors correspond to surfaces of sections made up of an infinite number of points that occupy a bounded domain of the cross-section. They may be chaotic or not. Hereafter, we plot Poincaré sections within a plane. Thus, we need an effective potential map to assess the geometry of the system under the trap. Here, the horizontal line represents regular motion, whereas chaotic motion is notified when a slope exists or the oscillation appears. 
	Finally, the spatial evolution of the wavefunction draws a picture of how the Boson evolves in time under several above considerations. 
A simple regular motion may easily be identified, while a complicated scheme denoted some chaotic behavior.

\section{Results and Discussion}

\subsection{BECs with constant two-body interaction alone}
\begin{figure}[!ht]%
	\centering\includegraphics[width=0.99\linewidth]{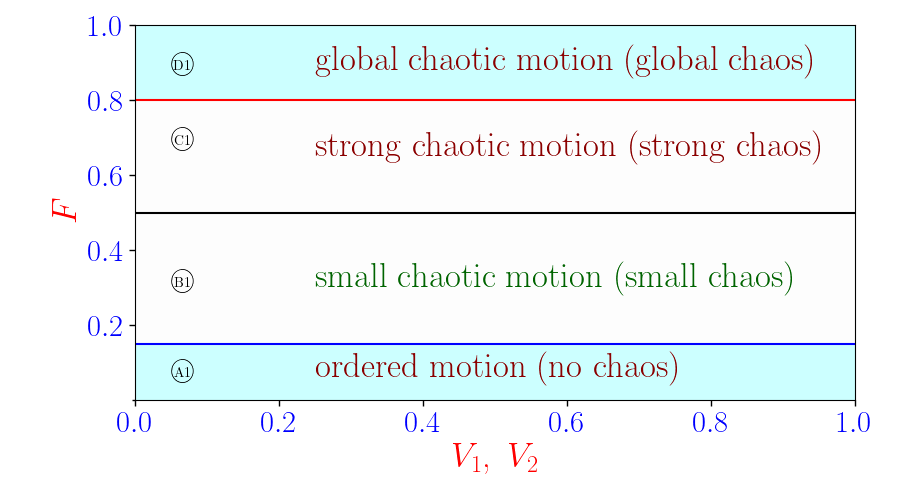}
\caption{(Color online) Chaotic domains for $g_0 = -1$. Parameters: $a = 1$, $b = 0$, $c = 0$, $d = 0$, $\chi_0 = 0$, $k_1 = 1$, $k_2 = 1$, $\mu = 0.0001$.}
\label{fdomain1}
	\label{fdomain1}%
\end{figure}

In this section, first, we consider the case-A of attractive BEC in the presence of a constant part of the two-body interaction alone, everything else is switched off. In Fig.\ref{fdomain1}, we show the different regimes for $V_1 (\text{or} V_2)$ Vs $F$ for different types of motions that occur in this case. For a weak value of the strength F, regular motion appear for all values of $V$. As  $F$ increases, different states of chaos are observed ranging from small to strong chaos, and later on global chaos emerges for very high $F$.

The observed chaotic dynamics in these regimes can be directly attributed to the presence of the lattice tilt, modeled by the linear term $Fx$ in the optical potential. This tilt introduces a constant energy gradient that breaks the spatial symmetry of the system, enabling nontrivial coupling between the condensate's modes, especially when combined with the system’s intrinsic nonlinearities. As the tilt strength $F$ increases, nonlinear resonances emerge, leading to instability and chaotic motion. Numerical indicators, including the Lyapunov exponents and Poincaré sections, confirm that even small values of $F$ can destabilize regular trajectories and give rise to small chaos, which eventually develops into strong and global chaos with further increases in tilt. This underlines the tilt strength $F$ as a key control parameter driving the transition to chaos in the condensate dynamics.

Next, we verify this occurrence through the four different tools afore-mentioned, by picking points within different regions. The $1^{st},2^{nd},3^{rd}$  and $4^{th}$ columns in Fig.\ref{fcase1} are for trajectory plots, Poincaré surface of section (SOS) plots, potential plots, and spacial evaluation plots, respectively, while $1^{st},2^{nd},3^{rd} and 4^{th}$ lines correspond to dynamics for the values picked from A1, B1, C1, and D1, respectively.

\begin{figure*}[!ht]
\centering\includegraphics[width=0.95\linewidth]{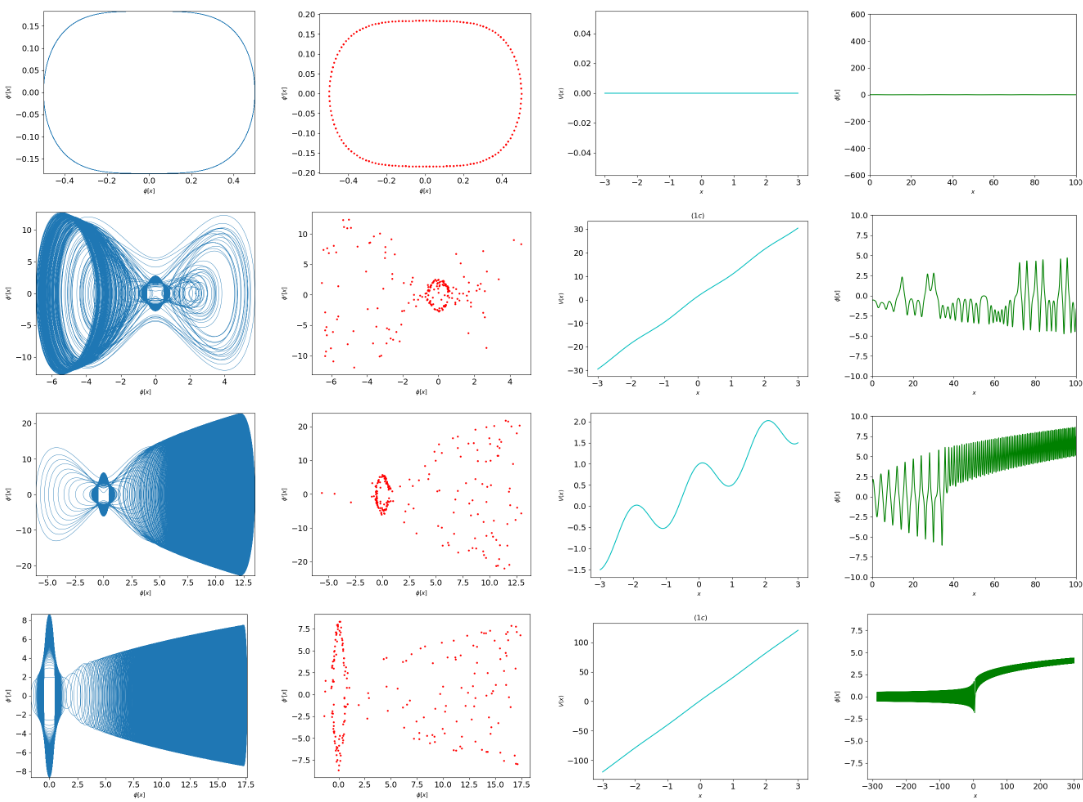}
\caption{Results for parameter sets A1, B1, C1, and D1 from Fig~\ref{fdomain1}. Rows from top to bottom correspond to A1–D1.}
\label{fcase1}
\end{figure*}

We investigate the regular and chaotic solutions of the BEC system with an attractive interaction in different parameter regions where the strength of field $F$ has varied from $F=0.0$ to $1.0$. We started our numerical analysis with a very weak field value $F=0.0$ and observed that the trajectory is still regular, showing the periodic behavior of the system. This is depicted in the first line of Fig.\ref{fcase1} for a point picked in region A1 of Fig.\ref{fdomain1}. All indicators seem to agree according to the pictures on the first line. When we gradually increased the value of $F$, we observed not-so-erratic irregular trajectories, a signature of chaos in the second line for point in domain B1 representing $F=0.2$ to $0.5$, where a chaotic motion starts. When we further increase the value of $F$, the results show that there is no longer any regular motion, and as the value of $F$ is increased strong chaos has emerged in the $3^rd$ line. All indicators exhibit pictures that traduce the chaotic behavior of the system. 
	In the final case which corresponds to very strong value $F$, there is no regular structure, and classical motion has been completely dominated by global Chaos (see last/bottom line) for a mode picked within domain D1.

\subsection{BECs with constant part of both two- and three-body interactions}

In the second step,  we consider the case-B of attractive BEC with the presence of a constant part for both the two-body and three-body interactions. In Fig.\ref{fdomain2}, we show the different regimes for $V_1 (\text{or} V_2)$ Vs $F$ for different types of chaoses that occur in this case. It appears that there is global chaos all over the plane. For every value of F, according to increasing values of V, always chaotic behavior was depicted. Selected modes A2, B2, C2, and D2 corresponding to previous domains where different types of motion were depicted, now stand for a single one, global chaos. This is further tested through the four different indicators presented earlier. The $1^{st},2^{nd},3^{rd},4^{th}$ columns in Fig.\ref{fdomain2} are for trajectory plots, Poincaré surface of section (SOS) plots, potential plots, and spacial evaluation plots, respectively, while $1^{st},2^{nd},3^{rd}$ and $4^{th}$ lines correspond to dynamics for the values picked from A2, B2, C2, and D2, respectively.

\begin{figure}[!ht]%
\centering\includegraphics[width=0.99\linewidth]{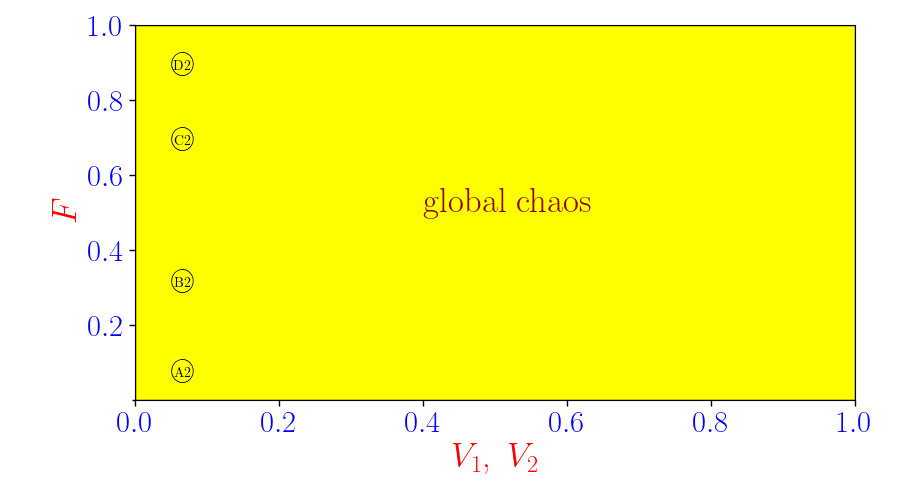}
\caption{(Color online) Chaotic domains for $g_0 = -1$ and $\chi_0 = -1$. Parameters: $a = 1$, $b = 0$, $c = 1$, $d = 0$, $k_1 = 1$, $k_2 = 1$, $\mu = 0.0001$.}
\label{fdomain2}%
\end{figure}

\begin{figure*}[!ht]
\centering\includegraphics[width=0.95\linewidth]{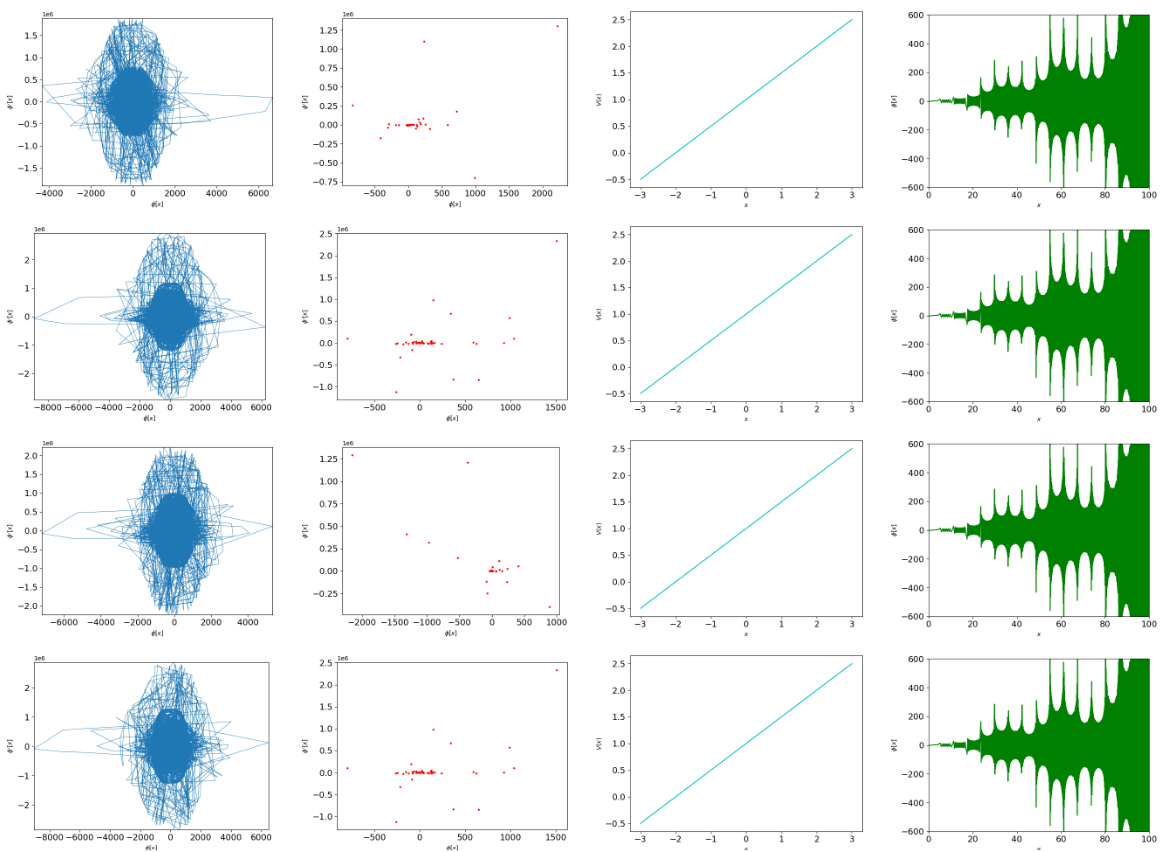}
\caption{(Color online) Results for parameter sets A2, B2, C2, and D2 from Fig.~\ref{fdomain2}. Each row (top to bottom) corresponds to A2, B2, C2, and D2, respectively, showing their dynamical behavior.}
\label{fcase2}
\end{figure*}

From all numeric indicators, we observe that the Poincaré surfaces of sections exhibit islands of points, showing how chaotic the dynamics of the BEC, Trajectory, and potential plots exhibit insight into complicated unpredictable behavior. The plots in Fig.\ref{fcase2} thus confirm that any selected mode in domains A2, B2, C2, and D2 trains the system into global chaotic motion.

With these selected values of parameters and the trapping potential, attractive BEC dynamics become globally chaotic when a constant three-body interaction term is taken into account. The regular motion could not survive when considering the second neighbor interaction term. This was also the case for repulsive BEC (not shown). For the said trapping potential, the BEC shares similarity with the modulational instability as earlier conjectured by ref.\cite{Lekeufack2013}.

\subsection{BECs with constant and oscillatory parts of two-body interaction}
\begin{figure}[!ht]%
\centering\includegraphics[width=0.99\linewidth]{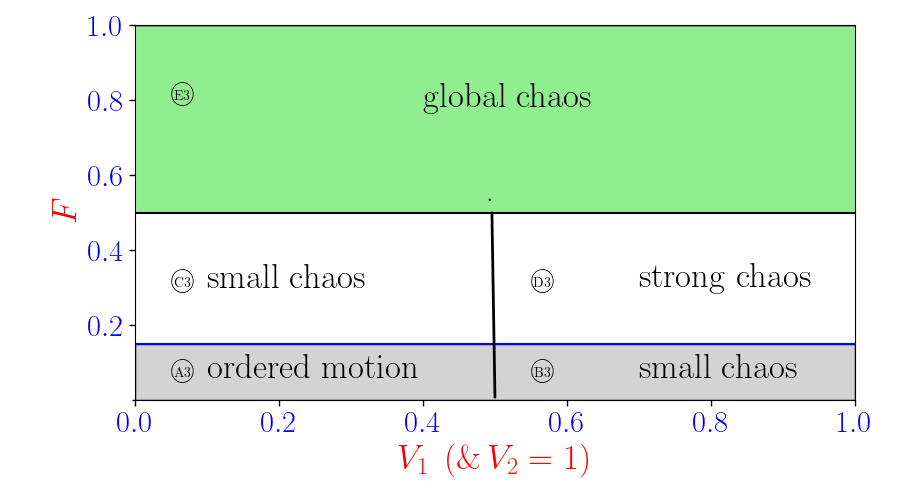}
\caption{(Color online) Chaotic domains for $g_0 = -1$ with $a = 1$, $b = 1$, $c = 0$, $d = 0$. Other parameters: $\chi_0 = 0$, $k_1 = 1$, $k_2 = 1$, $\mu = 0.0001$.}
\label{fdomain3}%
\end{figure}
\begin{figure*}[!ht]
\centering\includegraphics[width=0.95\linewidth]{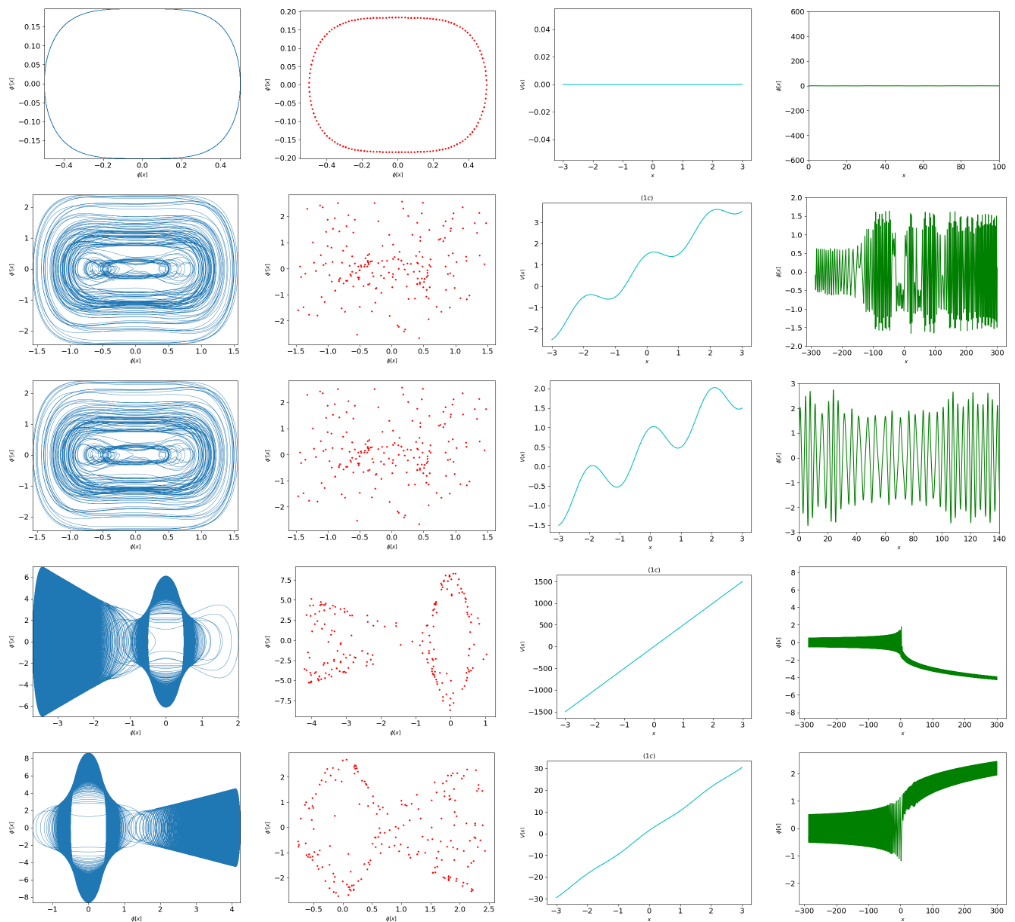}
\caption{Results for parameter sets A3, B3, C3, D3, and E3 from Fig.~\ref{fdomain3}. Rows from top to bottom correspond to A3–E3.}
\label{fcase3}
\end{figure*}

In case-C, we switch off the three-body interaction and activate both the AC and DC parts of the two-body interactions. As compared to Fig.\ref{fdomain1}, the plots show how former domains of strong chaos have turned to global chaos, offering larger chances for global chaos to occur due to the AC part of two-body interactions. Strong chaos now appears for the second half domain of former small chaos, for $V$ varying from $0.5$ to $1.0$. Small chaos that previously existed for all $V$ and for $F$ between the range $0.15$ - $0.5$ has now split into two subdomains i.e. first, for the same range of $F$ with $V$ varying from $0.0$ to $0.5$ and second, in the second half domain of previously ordered motion for the same range of $F$ but with V varying from $0.5$ to $1.0$. Finally, the domain of ordered motion is shrunk and reduced to range $0.0$ to $0.5$ of V. This is verified by numerical results in Fig.\ref{fcase3} where $1^{st},2^{nd},3^{rd},4^{th}$ and $5^{th}$  lines are for the values picked in regions A3, B3, C3, D3 and E3, respectively, as presented in Fig.\ref{fdomain3}. The first line exhibits regular ordered behavior as the close circle smooth curves and straight lines are obtained by our numerical indicators. Next, second and third lines represent small chaos, for points picked in domains B3 and C3. At last, strong and global chaoses are evidenced in the last two lines respectively where one observed a sea of points together with complicated unexpected behaviours.

The consideration of the oscillatory part of two-body interaction has annihilated the impact of three-body, by restoring all possible dynamics as in Fig.\ref{fdomain1} but providing more chances for global chaos to emerge for high values of $F$. In fact, the AC component of the  attractive two-body has modified the domain of various chaoses restricting regular motion to smaller $V$. So for attractive BEC to exhibit regular behavior with AC+DC, the trapping must be weak. Otherwise, for all $V$ with high $F$, many chances are there to threaten the system  into global chaos.

\subsection{BECs with both constant and oscillatory parts of both two- and three-body interactions}
\begin{figure}[!ht]%
\centering\includegraphics[width=0.99\linewidth]{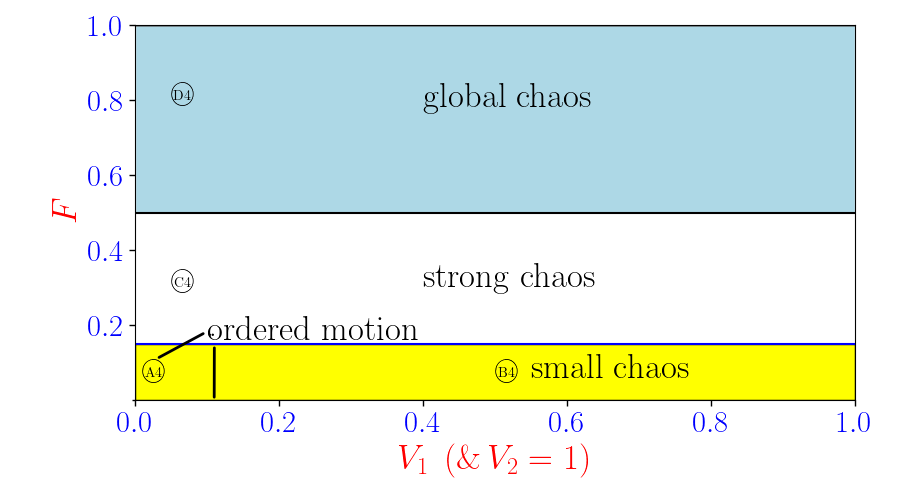}
\caption{(Color online) Chaotic domains for $g_0 = -1$, $\chi_0 = -1$ with $a = b = c = d = 1$. Parameters: $k_1 = 1$, $k_2 = 1$, $\mu = 0.0001$.}
\label{fdomain4}%
\end{figure}

In case-D, we consider both two- and three-body interactions with both AC and DC parts of the interaction. As compared to the case of AC+DC two-body interaction~Fig.\ref{fdomain3}, the picture in Fig.\ref{fdomain4} shows that global chaos due to AC+DC three-body is maintained, while strong chaos has expanded within the same range of $F$ but now for all values of V. So the AC of three-body has transformed small chaos into strong chaos. Also expanded is the domain of small chaos that now belongs to the range of V from $0.1$ to $1.0$, hence shrinking the bandwidth of ordered motion (values of V in the range $0.0$ to $0.1$). This is further verified by numerical results in Fig.\ref{fcase4} where $1^{st},2^{nd},3^{rd},4^{th}$  lines are for values picked in region A4, B4, C4 and D4  respectively. All indicators then confirmed the predictions.

The AC component of three-body interaction has removed the global chaos of  Fig.\ref{fdomain2} caused by its DC component on the constant two-body interaction term. This offered few possibilities for the BEC  to follow the regular motion. A system with AC+DC two- and three-body interactions is very chaos-sensitive as it threatens regular motion with large V into small chaos while upgrading small chaos into strong chaos (see Fig.\ref{fdomain4}). However, it has no effect on global chaos.

\begin{figure*}[!ht]
\centering\includegraphics[width=0.95\linewidth]{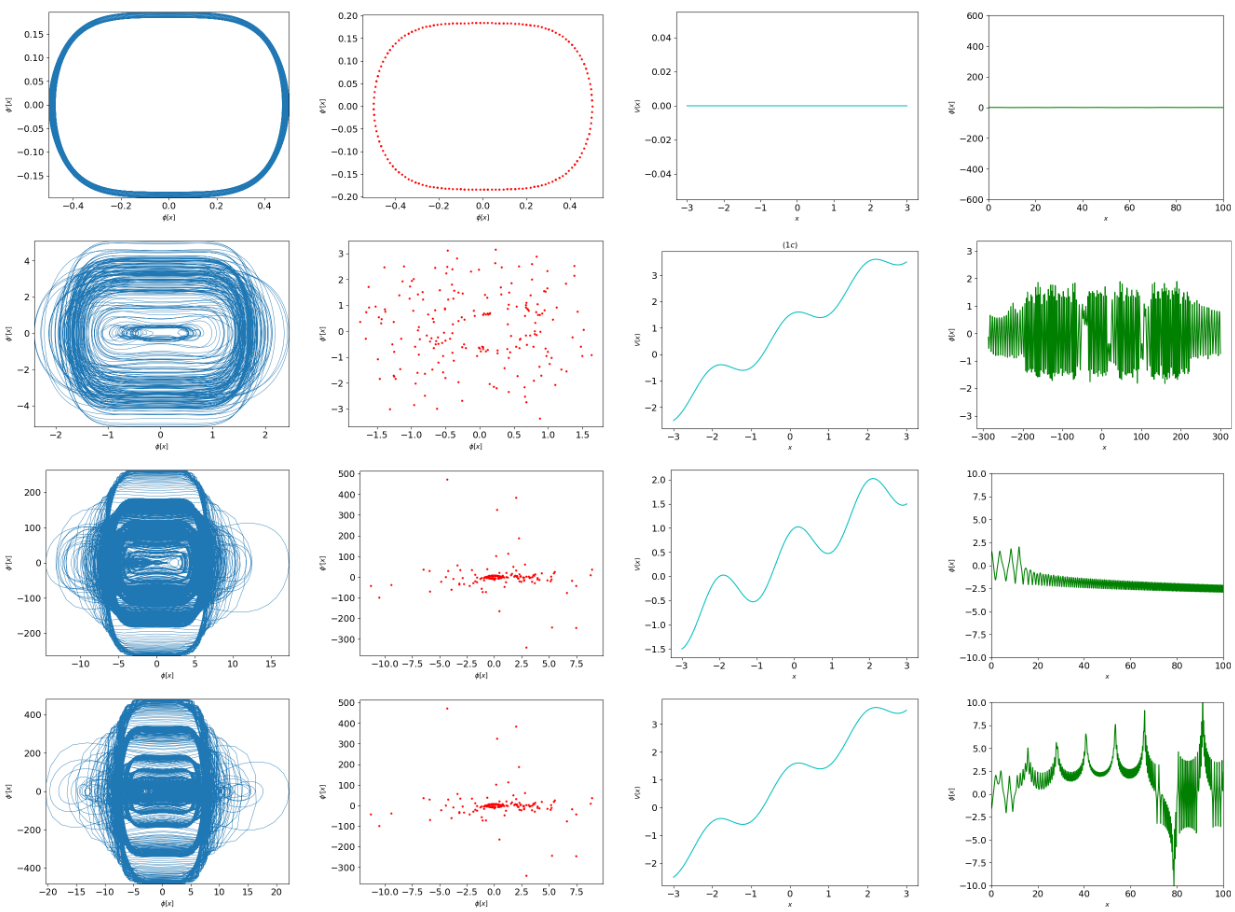}
\caption{Results for parameter sets A4, B4, C4, and D4 from Fig.~\ref{fdomain4}. Rows from top to bottom correspond to A4–D4.}
\label{fcase4}
\end{figure*}

\section{Lyapunov exponent}
The Lyapunov exponent plays a significant role in understanding the dynamics of BECs, especially when studying nonlinear phenomena such as quantum turbulence, vortex dynamics, and chaos in mean-field models like the GP equation.

	In general, the Lyapunov exponent is a numerical too commonly used in dynamical systems to quantify the rate of separation of infinitesimally close trajectories. 
It indicates the sensitivity of a system to initial conditions, a key characteristic of chaos. A system with at least one positive LE is considered chaotic, as small differences in initial conditions grow exponentially over time.
	The separation between two initially nearby trajectories was initialized at $\delta_0 \sim 10^{-6}$ and was renormalized periodically to avoid numerical overflow during the Lyapunov exponent computation.

\begin{figure}[!ht]
	\begin{center}
		\includegraphics[width=0.5\textwidth]{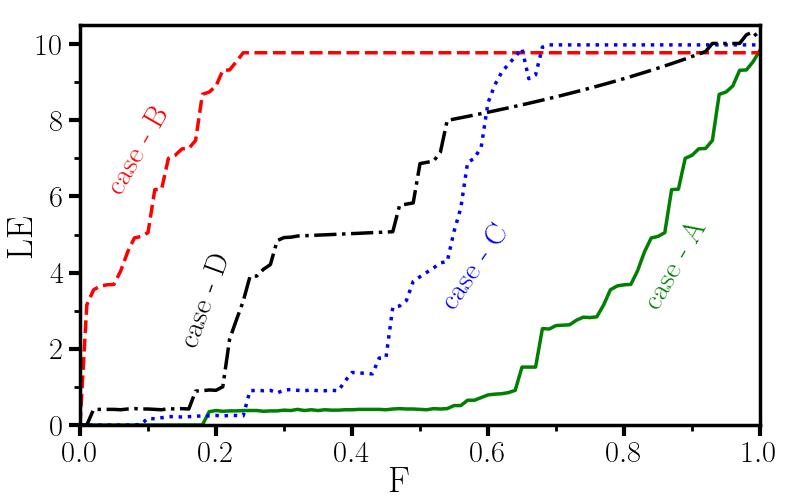}
	\end{center}
\caption{Lyapunov exponent results for cases A–D from Figs.~\ref{fdomain1}–\ref{fdomain4}: A (solid green), B (dashed red), C (dotted blue), and D (dash-dotted black).}
\label{flyapunov}
\end{figure}

This work involved calculating Lyapunov exponents over a range of $F$ values from $0.000$ to $1.000$ for four distinct cases, each covering different weak ranges. \\

{\bf From the solid green curve of case-A:}
\begin{enumerate}
	\item[{*}]For values of $F$ ranging from 0.001 to 0.180, the maximum Lyapunov exponent consistently indicated ordered motion, suggesting that the system remained primarily regular throughout this parameter range.
	
	\item[{*}] For values of $F$ ranging from 0.190 to 0.590, the maximum Lyapunov exponent transitioned to small positive values between 0.35 and 0.72, indicating minor chaotic behavior. However, the chaotic regions occupied only a minimal fraction of the phase-space volume across various initial conditions.
	
	\item[{*}] For values of $F$ ranging from 0.600 to 0.800, the maximum Lyapunov exponent increased significantly to a range between 0.82 and 3.60, marking the onset of pronounced chaotic dynamics.
	
	\item[{*}] For values of $F$ ranging from 0.800 to 1.000, the maximum Lyapunov exponent increased significantly to approximately 9.8, indicating widespread chaotic behavior throughout the entire system.
\end{enumerate}

{\bf From the dashed red curve of case-B:}

\begin{enumerate}
	\item[{*}] For values of $F$ spanning from 0.001 to 1.000, the maximum Lyapunov exponent consistently remained at a high positive value of approximately 9.72, indicating that classical motion was completely overwhelmed by global chaos, with no discernible regular structure.
\end{enumerate}

{\bf From the dotted blue curve of case-C:}

\begin{enumerate}
	\item[{*}] For values of $F$ ranging from 0.001 to 0.080, the maximum Lyapunov exponent was consistent with periodic motion, indicating that the system exhibited regular and ordered behavior.
	
	\item[{*}] For values of $F$ ranging from 0.090 to 0.230, the maximum Lyapunov exponent fluctuated between 0.12 and 0.24, indicating mild chaotic behavior that persisted across various initial conditions.
	
	\item[{*}] As $F$ increased from 0.240 to 0.520, the maximum Lyapunov exponent ranged between 0.91 and 4.21, confirming robust chaotic behavior.
	
	\item[{*}] As $F$ increased from 0.530 to 1.000, the maximum Lyapunov exponent reached approximately 9.97, marking widespread global chaos across all parts of the system.
\end{enumerate}

{\bf From the dash-dotted black curve of case-D:}

\begin{enumerate}
	\item[{*}] For values of $F$ ranging from $0.000$ to $0.001$, the maximum Lyapunov exponent was consistent with ordered motion, indicating that the system exhibited regular dynamics.
	\item[{*}] As $F$ increased from $0.210$ to $0.580$, the maximum Lyapunov exponent ranged between 1.05 and 7.09, signaling robust chaotic behavior.
	\item[{*}] For values of $F$ ranging from $0.600$ to $1.000$, the maximum Lyapunov exponent reached approximately $10.8$, indicating fully developed global chaotic dynamics throughout the system.    
\end{enumerate}    
\color{black}   

We have observed sensitivity to initial conditions, a key characteristic of chaotic dynamics, using the Lyapunov exponent. As the field strength increases from zero to higher values, regular regions emerge within chaotic domains, indicating a transition from small chaos to strong chaos and eventually global chaos. The maximal Lyapunov exponent analysis further reveals that the strength of the magnetic field directly influences the presence of chaos, with stronger fields leading to more pronounced chaotic behavior.

\section{Conclusions}\label{sec:dis}

In this paper, we have studied the dynamics of BEC subjected to optical tilted potential, in the consideration where AC and DC components of first and second neighbors are present in the description of the dynamics. It is seen that several management of the components for the scattering length leads to the occurrence of ordered motion, small, strong, and global chaoses in four distinct cases:
When the solely constant part of the two-body interaction is considered, there are chances of having regular motion, for the weak magnitude of strength F; while for increasing values, more and more chaotic behavior emerges until large $F$ with global chaos.
With the activation of the constant three-body interaction term, the systems lose all chances of moving in a regular way. In fact, even former regular motion from weak values of $F$ has now turned to global chaotic behavior.
Switching off the DC component of three-body and rather considering the oscillatory part of two-body interaction has modified the domain of various chaoses restricting regular motion to smaller $V$ (bandwidth of ordered motion) while extending that of global chaos. In fact, the AC component of the two-body interaction has provided more chances for global chaos to emerge in the second half plane of $F$. So, for attractive BEC to exhibit regular behavior with AC+DC, the trapping must be weak. Otherwise, for all $V$ with high $F$, many chances are there to threaten the system into global chaos. A system with both constant and oscillatory components of two- and three-body interactions is very chaos-sensitive as it threatens regular motion with large $V$ into small chaos while upgrading small chaos into strong one. However, it has no effect on global chaos. Few chances are given to such BEC systems to exhibit regular behavior.
Overall, the present  study shows how chaos is dominating the dynamics of BEC systems when particular managements of AC and DC components of two- and three-body interaction are taken into account. This provides some explanations on how sensitive is the dynamics of trapped BEC in optical tilted potential. This makes us ask the question whether or not these chaoses are connected to the control of BEC's stabilization process. In both theoretical and experimental fields, some applications of these results are in order of prevention over forthcoming behavior of such systems like modulational instability.

	A discussion point hereafter addressed needs careful attention of the researchers in condensed matter physics. The transitions between chaotic and ordered dynamics has a profound impact on the quantum features of BECs, especially in the presence of a tilted optical lattice and tunable nonlinear interactions. This impact is clearly observed through phase-space trajectories, wavefunction evolution, and Lyapunov exponent analysis.
	Ordered dynamics correspond to regular and periodic solutions of the Gross-Pitaevskii equation, supporting coherent quantum structures like solitons and stable density profiles. These features make ordered regimes ideal for applications in quantum sensing, atomic interferometry, and coherent matter-wave engineering \cite{ZhiXiaa2010}. For instance, our simulations in parameter regions A1 and A3 (see Figs.~\ref{fcase1} and \ref{fcase3}) demonstrate well-behaved, predictable wavefunction evolution, as indicated by smooth phase portraits and Poincaré sections.
	In contrast, chaotic dynamics—triggered by increased tilt strength $F$ and the inclusion of oscillatory (AC) or three-body interaction terms—lead to irregular, fragmented wavefunctions with high sensitivity to initial conditions. As shown in Figs.~\ref{fcase2} and \ref{fcase4}, these dynamics destroy the condensate's quantum coherence (quantum transport) and can result in modulational instability \cite{Lekeufack2013}, intricate patterns, or collapse. In addition, when coupled to environment, chaotic behavior of BEC may also lead to entanglement degradation (loss of correlation), and thus, affecting their control \cite{Furuya1998}. Such behavior underlines the utility of the BEC in precision quantum applications. The Lyapunov exponent, a classical indicator of chaos, serves here as a semi-quantitative tool to measure this transition. As shown in Fig.~\ref{flyapunov}, a positive maximal Lyapunov exponent corresponds to chaotic behavior, while near-zero values indicate ordered motion. The increase in Lyapunov exponent with $F$ confirms the emergence of global chaos, which significantly degrades the condensate’s quantum stability. Therefore, our study demonstrates that the chaotic or ordered nature of BEC dynamics critically influences their quantum properties. While ordered regimes preserve coherence and control, chaotic regimes induce unpredictability and quantum decoherence, posing challenges and opportunities for advanced quantum control strategies.

\section*{Acknowledgments}

	The authors are indebted to the referees for their positive and critical comments on the paper, which helped to improve the manuscript immensely.
S.S. acknowledges the Foundation for Research Support of the State of Sao Paulo (FAPESP) [Contracts 2020/02185-1, 2017/05660-0, 2024/04174-8, 2024/01533-7].

\end{document}